\documentclass[aps,prl,showpacs,onecolumn]{revtex4}
\usepackage{amsfonts}
\usepackage{amsmath}
\usepackage{amssymb}
\usepackage{epsfig,graphicx,amssymb}

\input{.tex}

\begin{document}

\title{Intermediate mass dileptons from the passage of jets and high energy photons through quark-gluon plasma}

\author{FU Yong-Ping$^{*}$,
 LI Yun-De
 }
\address{Department of Physics, Yunnan University, Kunming 650091,
China£»\\* E-mail:ynufyp@sina.cn}

\begin{abstract}
The production of the intermediate mass dileptons originating from
the annihilation and Compton scattering of the jets and high energy
photons (resolved photons) passing through the quark-gluon plasma is
calculated. The contribution of the dilepton yield due to the
jet-plasma and high energy photon-plasma interaction is pronounced
compared to the thermal and Drell-Yan dilepton spectrum at
intermediate mass. The ordinary spectrum of thermal and Drell-Yan
processes is enhanced by the jet and photon-plasma mechanism. The
numerical results match to the PHENIX data accurately in the
intermediate mass region for Au-Au 200 GeV/A collisions at RHIC.

\end{abstract}


\maketitle

The most important goal in the study of relativistic heavy ion
collisions is to probe the exact information of the quark-gluon
plasma (QGP). Since dileptons do not interact strongly, it is
relatively easy to probe the thermal dilepton information emitted
from the hot QGP in the short creating and cooling time
\cite{1,2,3,4,5,6,7,8,9,10,11,11.1}. However, so far no evident
experiments show that some information are exactly produced from the
QGP. For the theory of the phase transition, thermal dileptons are
dominant in the intermediate mass region between the $\phi$ and the
$J/\Psi$ vector meson, but the contribution of dileptons in this
mass region also can be explained by the decays of charmed mesons
\cite{19,20,21,22,23}.

Recently, the measurement of the dilepton continuum at Relativistic
Heavy Ion Collider (RHIC) energies was performed by the PHENIX
experiments for Au-Au 200 GeV/A collisions \cite{24}. The dilepton
yield in the low mass range between 0.2 and 0.8 GeV is enhanced by a
factor of 2$\sim$3 compared to the expectation from hadron decays.
In fact, such phenomena were found at Super Proton Synchrotron
(SPS), the enhancement of the dilepton yield at SPS was successfully
interpreted by the models of the mass dropping or melting in a hot
medium due to the chiral symmetry restoration, but such modifying
scenarios can not well explain the enhancement in Au-Au collisions
at RHIC \cite{24.1,12,13,14,15,16,17,18,26}. The imperfect modifying
models of hadron decays present other probable mechanisms to explain
the enhancement of the dilepton yield at low mass. Moreover, the
dilepton enhancement at RHIC is implied that such phenomena are
related strongly to the hot plasma scenario.

The contribution of thermal dileptons at the low mass is covered by
the cocktail of hadron decays due to the vector meson peaks is more
pronounced than the thermal spectrum. Then the thermal information
is only evident in the intermediate mass region. The dileptons
produced from jet-plasma and photon-plasma interactions contain a
thermal information coming from the quark-gluon plasma. Therefore
the dileptons produced from the passage of large transverse momentum
($P_{T}$) jets and photons passing through the hot medium are also
pronounced at intermediate mass. The dileptons produced in the large
$P_{T}$ jet-plasma and $\gamma$-plasma inelastic scatterings turns
into an important dilepton production source at intermediate mass.
This jet and photon-plasma conversion is absent in $p-p$ collisions.

The jet-dilepton conversion mechanism was discussed by Srivastava
before \cite{29.2,29.3}. However, it is the first time to rigorously
discuss the jet-dilepton and photon-dilepton conversions for the hot
medium system in this Letter.

In pQCD the transverse momentum of photons can arise by the hard
bremsstrahlung of high energy gluons which can be calculated
perturbatively if the momentum transfers are large. The perturbative
component of the large $P_{T}$ photons is generated by the following
subprocess: $q\bar{q}\rightarrow g \gamma$, $qg\rightarrow q \gamma$
and $qq\rightarrow (q\rightarrow q\gamma) q $ \cite{11.1,28,29}. The
idea of $\gamma$-plasma interaction is based on the QED Compton
cross section: $q_{th} \gamma \rightarrow q_{th}
(\gamma^{*}\rightarrow l\bar{l})$ when a high energy photon passing
through the quark-gluon plasma, where $q_{th}$ denote the thermal
partons in the hot plasma.

Furthermore the dilepton production associated with resolved
photon-plasma interaction is also discussed. Heisenberg uncertainty
principle allow a photon for a short time also to fluctuate into a
quark-antiquark pair. Therefore a high energy photon looks like
surrounded by a quark cloud, and can be interpreted that it has a
inner parton structure \cite{29.1}. When the photons emitted by the
hard collisions have large transverse momentum, the photons which
include inner parton structure are hadron-like, so a target thermal
parton in the medium reacts with the photon-parton in the processes
of $q_{\gamma}\bar{q}_{th}\rightarrow \gamma^{*}\rightarrow
l\bar{l}$, $q_{\gamma}\bar{q}_{th}\rightarrow g
(\gamma^{*}\rightarrow l\bar{l})$, $q_{\gamma}g_{th}\rightarrow q
(\gamma^{*}\rightarrow l\bar{l})$ and ($q_{th} g_{\gamma}\rightarrow
q \gamma^{*}$), where the photon-parton
$q_{\gamma}/\bar{q}_{\gamma}$ and $g_{\gamma}$ depends on the large
$P_{T}$ carried by the high energy photons.

In this paper, we rigorously derive the dileptons production rate of
the high energy jet-dilepton and photon-dilepton conversion. Let us
start with considering the yield of dileptons in the jet-plasma and
photon (resolved photon)-plasma interactions. Using kinetic theory
of the two body interaction $p_{1}p_{2(th)}\rightarrow
p_{3(l\bar{l})}p_{4}$, the production rate is
$R_{l\bar{l}}\propto\int d^{3}\tilde{p}_{1}\int
d^{3}\tilde{p}_{2}f(p_{1})f(p_{2})v_{12}\sigma$, where
$d^{3}\tilde{p}=d^{3}p/(2\pi)^{3}$ and the relative velocity is
$v_{12}=(p_{1}+ p_{2})^{2}/2p_{1}^{0} p_{2}^{0}$. In the limit
$P^{l\bar{l}}_{T}\rightarrow 0$ the momentum integration finally can
be written into a one dimension form
\begin{eqnarray}
\frac{dR_{l\bar{l}}}{dM^{2}}\label{eq1}=\frac{6}{(2\pi)^{4}}\int
dP_{T}f_{jet/q_{\gamma}}(P_{T})TM^{2}e^{-\frac{M^{2}}{4P_{T}T}}\sigma_{DY},
\end{eqnarray}
where $f_{jet/q_{\gamma}}$ is the distribution of large $P_{T}$ jets
and the distribution of partons in a high energy photon,
respectively. Note that the gluon jets and real photons will
contribute only at high order. Here
$\sigma_{DY}=K\frac{4\pi\alpha^{2}e_{q}^{2}}{9M^{2}}$ is the
standard cross section of Drell-Yan process with a $K$ factor
$K=1+2.09\alpha_{s}$. If the distribution is replaced by the thermal
Bltzmann distribution $e^{-P_{T}Coshy/T}$, one can derive the
standard rate of the thermal dileptons production as \cite{1}
$dR_{l\bar{l}}/dM^{2}=\frac{6\sigma_{DY}}{(2\pi)4}M^{3}TK_{1}(\frac{M}{T})$.

The phase-space distribution of jets and high energy photons
(resolved photons) is as follows
\begin{eqnarray}
f_{jet/q_{\gamma}}(p)=\frac{2(2\pi)^{3}}{g_{q/\gamma}V_{ch}
P_{T}coshy}\frac{dN _{jet/q_{\gamma}}}{d^{2}P_{T}dy}\delta(\eta-y),
\end{eqnarray}
where $g_{q/\gamma}$ is the spin(polarization) and color degeneracy
of a quark and  a photon, respectively. Here $g_{q}=6$ and
$g_{\gamma}=2$. $\eta$ is the space-time rapidity of the hot system.
$V_{ch}$ is the system volume. One can treat the production rate of
the jets and photons by scaling the results for the cross section of
Nucleon-Nucleon collisions with the the nuclear thickness for a
head-on collision in the form $
dN/d^{2}P_{T}dy=T_{AA}d\sigma/d^{2}P_{T}dy$, where the nuclear
thickness for zero impact parameter is $9A^{2}/8\pi R^{2}_{\bot}$,
$R_{\perp}$ is the radius of the fireball. we take the initial
radius of the QGP as $R_{\perp}$= 4$\sim$8 fm for RHIC Au-Au
$\sqrt{S_{NN}}$=200GeV collisions \cite{29.3}.

We choose the accelerating expanding volume of the cylindrical
hydro-type as
$V_{ch}(\tau)=2(z_{0}+v_{z}\tau+\frac{1}{2}a_{z}\tau^{2})\pi(R_{0}+\frac{1}{2}a_{\perp}\tau^{2})^{2}$
\cite{3}. The value of $z_{0}$ equals to the QGP formation time
$\tau_{0}$. After expanding the terms of $V_{ch}(\tau)$, a simple
form of the system volume can be deduced as $
V_{ch}(\tau)=V_{0}+\sigma\tau+a_{V}\tau^{2}+\mathcal
{O}(\tau^{3},\tau^{4},\tau^{5},\tau^{6}) $, where the parameters
$V_{0}=2\pi R^{2}_{0}z_{0}$, $\sigma=2\pi v_{z}R^{2}_{0}$ and
$a_{V}=\pi R^{2}_{0}a_{z}+2\pi R_{0}z_{0}a_{\perp}$. The authors in
Ref. \cite{1} take the value of the transverse area as $\sigma\sim$
100 $fm^{2}$ for $T_{0}\sim$ 200-300 MeV. If we choose the radius of
the transverse area as $R_{QGP}\sim$ 4 fm, one can immediately find
that the longitudinal velocity $|v_{z}|\sim$ 1. The values of the
parameters $a_{z}$ and $a_{\perp}$ in the accelerating terms are
relatively smaller than the value of $v_{z}$, which means that the
term $\mathcal {O}(\tau^{3},\tau^{4},\tau^{5},\tau^{6})$ is
negligible. These accelerations are adjusted to the final conditions
of flow velocities.

We now turn our attention to the production rate of the jets and
photons. The cross section of high energy jets ($qq \rightarrow qq$,
$qg \rightarrow qg$, $q\bar{q}\rightarrow gg$, $gg\rightarrow
q\bar{q}$ and $gg \rightarrow gg$) is given by \cite{29}
\begin{eqnarray}
\frac{d\sigma_{jet}}{d^{2}P_{T}dy}\label{eq1}&=&\int^{1}_{x_{a}^{min}}
dx_{a}G_{A/a}(x_{a},Q^{2})G_{B/b}(x_{b},Q^{2})\nonumber\\[1mm]
&&\times \frac{x_{a}x_{b}}{x_{a}-x_{1}}
\frac{1}{\pi}\frac{d\hat{\sigma}_{ab\rightarrow jet}}{d\hat{t}},
\end{eqnarray}
we choose the parton distribution $G_{N/n}(x,Q^{2})$ of the nucleon
from GRV \cite{28} in the form $
G_{N/n}(x,Q^{2})=R(x,Q^{2},A)[ZP(x,Q^{2})+(A-Z)N(x,Q^{2})]/A$, where
$R(x,Q^{2},A)$ is the nuclear shadowing factor \cite{30}, $Z$ is the
proton number of the nucleus and $A$ is the nucleon number.
$P(x,Q^{2})$ is the proton distribution, and $N(x,Q^{2})$ is the
neutron distribution. Since protons and neutrons have different
distribution of up and down valence quarks, the effect of the
iso-spin can be represented by the sum of the proton and neutron
distribution. The treatment that without the shadowing and iso-spin
effect will overestimate the dilepton contribution in the high
$P_{T}$ region \cite{26}. The best scale of the transverse momentum
is $Q^{2}=p^{2}_{a,bT}=4P_{T}^{2}$ in the $Q^{2}$ dependent QCD
structure functions $G_{N/n}(x,Q^{2})$. One can find the
photon-parton distribution function from M. Gl$ \ddot{\mathrm{u}}$ck
et al. We have taken the $Q^{2}_{\gamma}=q^{2}_{T}=p^{2}_{a,bT}$
\cite{28}.

The large $P_{T}$ photons emitted by the subprocesses
$q\bar{q}\rightarrow g \gamma$ (ann.) and $qg\rightarrow q \gamma$
(Com.) in the heavy ion collisions satisfy the cross section
$(A+B\rightarrow \gamma+X)$ in the following \cite{11.1}
\begin{eqnarray}
\frac{d\sigma_{dir.-\gamma}}{d^{2}P_{T}dy}\label{eq1}&=&\int^{1}_{x_{a}^{min}}
dx_{a}G_{A/a}(x_{a},Q^{2})G_{B/b}(x_{b},Q^{2})\nonumber\\[1mm]
&&\times \frac{x_{a}x_{b}}{x_{a}-x_{1}}
\frac{1}{\pi}\frac{d\hat{\sigma}_{Com., ann.}}{d\hat{t}},
\end{eqnarray}
where the minimum volume of the momentum fraction are
$x_{a}^{min}=x_{1}/(1-x_{2})$, and the fraction of nucleon $B$ is
$x_{b}=x_{a}x_{2}/(x_{a}-x_{1})$, here
$x_{1}=\frac{P_{T}}{\sqrt{S_{NN}}}e^{y}$,
$x_{2}=\frac{P_{T}}{\sqrt{S_{NN}}}e^{-y}$.

\begin{figure} [h]
\includegraphics[angle=0,scale=0.8]{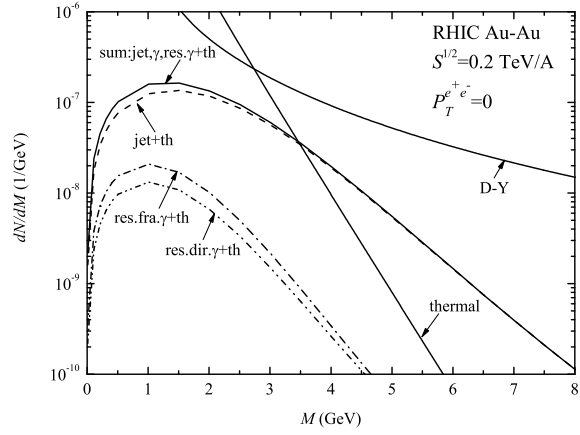}  
\vspace*{0.0cm} \caption{ The dileptons produced from jet-plasma and
$\gamma$(res. $\gamma$)-plasma interactions at RHIC for Au-Au
$\sqrt{S_{NN}}$=200 GeV collisions. The contribution of the
$\gamma$-dilepton conversion is included into the $K$ factor of the
jet-dilepton conversions.}
\end{figure}

The cross section of photon fragmentation processes $qq\rightarrow
(q \rightarrow q\gamma)q$ is given by the following \cite{11.1}
\begin{eqnarray}
\frac{d\sigma_{fra.-\gamma}}{d^{2}P_{T}dy}\label{eq1}&=&\int^{1}_{x_{a}^{min}}
dx_{a}\int^{1}_{x_{b}^{min}} dx_{b}G_{A/a}(x_{a},Q^{2})\nonumber\\[1mm]
&&\times G_{B/b}(x_{b},Q^{2})D_{q}^{\gamma}(z,Q^{2})\nonumber\\[1mm]
&&\times \frac{1}{z\pi}\frac{d\hat{\sigma}_{qq\rightarrow
qq}}{d\hat{t}},
\end{eqnarray} 
where $D_{q}^{\gamma}(z,Q^{2})$ is the photon fragmentation
function. The minimum volume of the fraction for parton $b$ is
$x_{b}^{min}=x_{a}x_{2}/(x_{a}-x_{1})$, and the fraction
$z=x_{1}/x_{a}+x_{2}/x_{b}$. One can see the cross sections for the
subprocesses of $q\bar{q}\rightarrow g \gamma$, $qg\rightarrow q
\gamma$ and $qq\rightarrow q q $ in the Ref. \cite{29}.

The yield of photon-partons relevant to the Compton, annihilation
and fragmentation processes are derived in the following
\begin{eqnarray}
\frac{d\sigma_{res.dir.-\gamma}}{d^{2}P_{T}dy}\label{eq1}&=&\int^{1}_{x_{a}^{min}}
dx_{a}\int^{1}_{x_{b}^{min}}
dx_{b}G_{A/a}(x_{a},Q^{2})\nonumber\\[1mm]
&&\times
G_{B/b}(x_{b},Q^{2})G_{\gamma/q_{\gamma}(g_{\gamma})}(z,Q_{\gamma}^{2})\nonumber\\[1mm]
&&\times \frac{1}{z\pi}\frac{d\hat{\sigma}_{Com., ann.}}{d\hat{t}},
\end{eqnarray}
and
\begin{eqnarray}
\frac{d\sigma_{res.fra.-\gamma}}{d^{2}P_{T}dy}\label{eq1}&=&\int^{1}_{x_{a}^{min}}
dx_{a}\int^{1}_{x_{b}^{min}} dx_{b}\int^{1}_{z_{1}^{min}}
dz_{1}\nonumber\\[1mm]
&&\times G_{A/a}(x_{a},Q^{2})G_{B/b}(x_{b},Q^{2})\nonumber\\[1mm]
&&\times D_{q}^{\gamma}(z_{1},Q^{2})G_{\gamma/q_{\gamma}(g_{\gamma})}(z_{2},Q_{\gamma}^{2})\nonumber\\[1mm]
&&\times
\frac{1}{z_{1}^{2}z_{2}\pi}\frac{d\hat{\sigma}_{qq\rightarrow
qq}}{d\hat{t}},
\end{eqnarray}
where $G_{\gamma/q_{\gamma}(g_{\gamma})}(z,Q_{\gamma}^{2})$ is the
parton distribution of the resloved photons, here
$z^{min}_{1}=x_{1}/x_{a}+x_{2}/x_{b}$ and
$z_{2}=x_{1}/z_{1}x_{a}+x_{2}/z_{1}x_{b}$.

\begin{figure} [h]
\includegraphics[angle=0,scale=0.8]{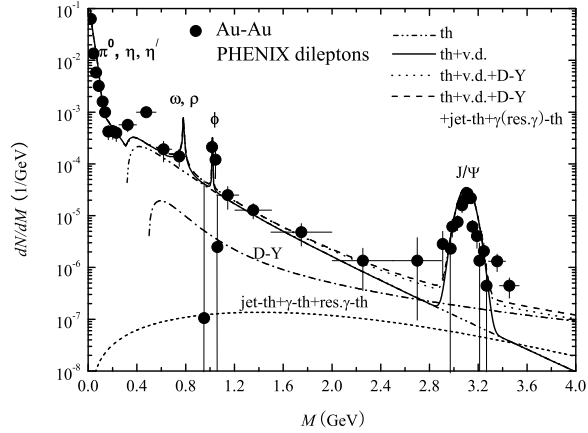}  
\vspace*{0.0cm} \caption{ The dileptons produced from jet-plasma and
$\gamma$(res. $\gamma$)-plasma interactions with PHENIX data for
Au-Au $\sqrt{S_{NN}}$=200 GeV collisions at intermediate mass. Data
from Ref. \cite{24}.}
\end{figure}

Since direct real photons are produced directly from the two body
kinetic reactions $qg\rightarrow q\gamma$ and $q\bar{q}\rightarrow
\gamma g$. The transverse momentum of direct real photons can arise
by the direct hard bremsstrahlung of high energy gluons. The large
$P_{T}$ carried by the real photons depends on the cross sections of
$ab\rightarrow c \gamma$ and $ab\rightarrow c (d\rightarrow
d\gamma)$ subprocesses. However, the yield of real photon-plasma may
be depressed by the mean cross sections $\sigma(q_{th} \gamma
\rightarrow q \gamma^{*})$ due to the lower coupling parameter
$\alpha^{3}$. In the integration of mean cross section
$\sigma=\int\frac{d\sigma}{dt}(P_{T})dt$, a divergence exists since
the Mandelstam term $1/\hat{t}$ in the cross section is divergent in
the limit $\hat{t}\rightarrow0$. After the virtual gluon
regularization \cite{11.1}, a finite results can be expressed as
$\sigma(q_{th}\gamma\rightarrow ql\bar{l}
)=\frac{\alpha}{\pi}\sigma_{DY}$ which can be contained into the $K$
factor of the jet-plasma processes. In the hot medium the
traditional yield of the Drell-Yan process
$q_{th}\bar{q}_{th}\rightarrow \gamma^{*}\rightarrow l\bar{l}$ does
not include the Mandelstam terms of $1/\hat{t}$, the processes will
not diverge in the infrared limit $P^{e^{+}e^{-}}_{T}\rightarrow 0$.
In fact, the energy spectrum of Drell-Yan process is longitudinal,
namely $E_{\gamma^{*}}^{2}=P_{L}^{2}+M^{2}$, where $P_{L}$ is the
longitudinal momentum of dileptons.

In Fig. 1 we plot the contributions for jet-dilepton and resolved
photon-dilepton conversion without the vacuum vector meson decays.
The value of the jets distribution $f_{jet}(p)$ is larger than the
resolved photons distribution due to the coupling parameter
($\alpha\alpha_{s}$) depresses the contribution of photon-partons.
Therefore the yield of the resolved photon-plasma reaction partly
modify  the jet-plasma yield. The contribution of the resolved
photon-dilepton conversion is almost 26\% of the sum for the
jet-plasma and $\gamma$(resolved $\gamma$)-plasma contribution from
0.5GeV-2.0GeV. The contribution of the real photon-plasma is
contained in the $K$ factor of jet-plasma processes. In Ref
\cite{29.2} the authors derived the jet-dilepton spectrum which is
parallel to the Drell-Yan yield at the RHIC energy. In this Letter
the jet-dilepton spectrum is depressed quickly with the increase of
$M^{2}$ due to the attenuation thermal function $e^{-M^{2}/4P_{T}T}$
in equation (1).

From Fig. 2 one can see that the spectrum of dileptons is enhanced
by the jet/$\gamma$(resolved $\gamma$)-plasma interaction mechanism
(dash line) in the intermediate mass region between 1.0 GeV to 2.8
GeV compared to the traditional spectrum. The dot line means the
contribution of the traditional thermal yield and Drell-Yan
processes, and the expectation from the meson vacuum decays(v. d.).
The numerical results match to the PHENIX data accurately at
intermediate mass. The enhancement at low mass is not considered in
this Letter, and the jet/$\gamma$(res.$\gamma$)-plasma interaction
is just a weak contribution at low mass.

As a conclusion, we rigorously derive the yield of dileptons for the
large $P_{T}$ jet-plasma and $\gamma$(resolved $\gamma$)-plasma
interaction mechanism. We find that the contribution of the
jet-plasma and $\gamma$(resolved $\gamma$)-plasma dileptons is
pronounced at intermediate mass, this mechanism satisfy the PHENIX
data at intermediate mass.

\begin{acknowledgments}
This work is supported by the National Natural Science Foundation of
China under Grant No: 10665003.
\end{acknowledgments}

\end{document}